\documentclass[12pt]{article} 
\usepackage{amsmath} 
\usepackage{amsfonts} 
\usepackage{amssymb} 
\usepackage{latexsym} 
\usepackage{graphicx} 
\usepackage[english]{babel} 
\input epsf.sty 
 
\newcommand{\be}{\begin{equation}} 
\newcommand{\ee}{\end{equation}} 
\newcommand{\ba}{\begin{array}} 
\newcommand{\ea}{\end{array}} 
\newcommand{\bqa}{\begin{eqnarray}} 
\newcommand{\eqa}{\end{eqnarray}} 
 
\begin{document}

\title{String Gas Cosmology and Non-Gaussianities} 
 \vspace{3mm} 
 
 \author{{Bin Chen$^{1}$, Yi Wang$^{2,3}$, Wei Xue$^{1}$, Robert Brandenberger$^{4}$}\\ 
{\small $^{1}$ Department of Physics, Peking University, Beijing 100871, P.R.China}\\ 
{\small $^{2}$ Institute of Theoretical Physics, Academia Sinica, 
Beijing 100080, P.R.China}\\ 
{\small $^{3}$  The Interdisciplinary Center for Theoretical Study,}\\ 
{\small University of Science and Technology of China (USTC), Hefei, 
Anhui 230027, P.R.China}\\ 
{\small $^{4}$ Physics Department, McGill University, Montreal, QC, 
H3A 2T8, Canada}} 
\date{}

\maketitle 
 
\begin{abstract} 
 
Recently it has been shown that string gas cosmology, an alternative 
model of the very early universe which does not involve a period of 
cosmological inflation, can give rise to an almost scale invariant 
spectrum of metric perturbations. Here we calculate the 
non-Gaussianities of the spectrum  of cosmological fluctuations in 
string gas cosmology, and find that these non-Gaussianities depend 
linearly on the wave number and that their amplitude depends 
sensitively on the string scale. If the string scale is at the TeV 
scale, string gas cosmology could lead to observable 
non-Gaussianities, if it is close to the Planck scale, then the 
non-Gaussianities on current cosmological scales are negligible.

\end{abstract} 

\noindent Preprint CAS-KITPC/ITP-023 
 
\newpage 
 
\section{Introduction} 
 
One of various alternative scenarios to inflation is string gas 
cosmology (SGC) \cite{BV,TV}. Unlike the usual inflationary models, 
SGC does not assume a quasi-de Sitter phase in which the universe 
undergoes an exponential growth. Instead, in SGC, there exists a 
quasi-static Hagedorn phase, during which thermal fluctuations of 
closed strings generate the density perturbations which seed the 
wrinkles in the cosmic microwave background (CMB) and the large 
scale structure of our universe today. In \cite{NBV,BNPV1}, it was 
shown that metric perturbations in SGC yield a scalar power spectrum 
consistent with current experiments, and predict a slight blue tilt 
to the spectrum of gravitational waves \cite{BNPV2}. 
 
SGC can be embedded naturally into string theory, which is the most 
promising candidate for quantum gravity. Potentially, SGC can 
overcome some of the conceptual problems of the inflationary 
scenario such as the singularity problem \cite{Borde} and the 
trans-Planckian problem for fluctuations \cite{Jerome}. 
 
SGC is based on taking into account the new degrees freedom 
(oscillatory and winding modes) and new symmetries (T-duality) of 
string theory. The oscillatory modes lead to a maximal temperature 
(Hagedorn temperature \cite{Hagedorn}) which a gas of strings can 
attain. In turn, the presence of a maximal temperature gives rise to 
the hope that in SGC one may avoid the cosmic singularity. Also, 
through the dynamical process of collision of string winding modes, 
our universe can naturally evolve from (9+1) or (10+1) compact 
dimensions to a space-time with (3+1) large dimensions where we live 
today \cite{BV,TV}. In short, SGC opens a new window to study the 
early universe. Various issues have been discussed in the literature 
(see \cite{RHBrev1,RHBrev2,BattWat} for reviews and references
to other work on string gas cosmology). 
In particular, string gas cosmology provides an elegant mechanism to 
stabilize most of the string moduli 
\cite{Watson,Patil1,Patil2,Edna,Watson2,Sugumi}, the one exception 
being the dilaton field. The dynamics of the compact dimensions 
relative to our three large spatial ones is studied in more detail 
in \cite{Mairi,Col,Danos}. 
 
In this paper, we would like to discuss the non-Gaussianities in 
string gas cosmology. Non-Gaussianity is one of the most important 
quantities which can be measured in upcoming experiments. 
Non-Gaussianity is characterized by amplitude, shape, sign and even 
running. Thus, it may contain a lot of information about the very 
early universe, and can be used to rule out models and set 
constraints on model building. 
 
In practise, the amount of non-Gaussianity is often described using 
the quantity $f_{\rm NL}$ \cite{NG}, which is of the order 
\begin{equation} 
  f_{\rm NL} \, \sim \, 
\frac{{\cal R}_g-{\cal R}}{{\cal R}_g^2-\langle {\cal 
R}_g^2\rangle}~, 
\end{equation} 
where $\cal R$ is the comoving curvature perturbation, and the 
subscript $g$ denotes its Gaussian part. 
 
In \cite{we}, we developed a method to calculate the 
non-Gaussianities of the fluctuations with thermal origin. Since in 
string gas cosmology matter in the early Hagedorn phase is dominated 
by a thermal gas of strings, the perturbations are of thermal 
origin. Hence, the techniques of \cite{we} can be used to 
study the non-Gaussianities in SGC and see under which conditions 
one obtains a value of $f_{NL}$ which is in the range of anticipated 
observations. 
 
This paper is organized as follows. In Section 2, we describe the 
evolution of the space-time background during the Hagedorn and 
radiation-dominated phases. In Section 3, we use thermal correlation 
functions to calculate the power spectrum of scalar metric 
perturbation and the non-Gaussianities. 
 
\section{Space-time Background in String Gas Cosmology} 
 
In string gas cosmology, the radiation phase of standard cosmology 
is preceded by the Hagedorn phase, a phase during which the 
temperature hovers near its maximal value, the Hagedorn temperature. 
Based on the symmetries of string theory, it is reasonable to assume 
that this phase is quasi-static in the sense that both the scale 
factor and the dilaton are constant. Einstein gravity is 
inconsistent with this behavior since it does not contain one of the 
key symmetries of string theory, T-duality. A better action to use 
is dilaton gravity, given by the effective action 
\be S \, = \, -\int d^{d+1}x 
\sqrt{-g}e^{-2\phi}[\Re+4g^{\mu\nu}\partial_\mu \phi \partial_\nu 
\phi]~, \ee 
where $\Re$ is the Ricci scalar in the string frame, $g$ is the 
determinant of the metric of the $d+1$ dimension space-time, and 
$\phi$ is the dilaton field. The topology of space is $T^d \times 
T^{9-d}$, where $d = 3$ is the number of the large spatial 
dimensions, and the other $9 - d$ spatial dimensions are stabilized 
at a microscopic scale. The above effective action only describes 
the expanding space-time. The conditions to use the above action are 
that the string coupling is small and the moduli of the extra 
compact small dimensions are stabilized. Similar to what is done in 
Friedmann cosmology, the early universe is assumed to be homogeneous 
and isotropic, given by the metric 
\be ds^2 \, = \, dt^2 - a(t)^2 d{\bf x}^2 ~ . \ee 
 
Through direct calculation, it follows that the variational 
equations of motion in the string frame can be expressed as 
\cite{TV}, 
\bqa -(d) {\dot \lambda}^2 + {\dot \varphi}^2 \, &=& \, e^{\varphi} 
E 
\label{E1} \\ 
{\ddot \lambda} - {\dot \varphi} {\dot \lambda} \, &=& \, 
{1 \over 2} e^{\varphi} P \label{E2} \\ 
{\ddot \varphi} - (d) {\dot \lambda}^2 \, &=& \, {1 \over 2} 
e^{\varphi} E \, , \label{E3} \eqa 
where $E$ and $P$ are the total energy and pressure in $(d+1)$ 
dimension space-time, and for convenience we introduced 
$\lambda=\ln[a(t)]$ as the logarithm of the scale factor, and 
rescaled the dilaton $\varphi= 2 \phi -d \lambda$.

The dynamics of space-time in SGC results by coupling a thermal gas 
of strings as matter (determining the energy and the pressure) to 
the above background. The new degrees of freedom and new symmetry of 
string theory lead to a cosmology which is rather different from 
what is obtained in standard cosmology. Let us follow the universe 
back in time. At large radii, all of the energy of the strings is in 
string modes which behave as usual radiation. It can be shown that 
in the case of a radiative equation of state the above background 
equations reduce to the usual Einstein equations, and the dilaton 
can be set to be constant. Thus, at late times SGC and Standard Big 
Bang cosmology predict the same evolution of the scale factor. 
However, at small radii the string energy will begin to flow into 
the winding modes. This, in turn, will lead to a reduction in the 
pressure. When the radius equals the string scale (or, more 
generally, when the energy density exceeds the string energy 
density), there is an equal amount of energy in the winding and 
momentum modes, and their contribution to the pressure cancels. It 
the total pressure vanishes, then, as follows immediately from 
(\ref{E2}), the scale factor tends to a constant. Thus, at high 
densities the universe hovers in a quasi-static Hagedorn phase. 
 
One set of relevant new degrees of freedom are the winding modes of 
the closed string, and the new symmetry is the T-duality. To 
illustrate these new features, let us consider string theory on a 
circle $T^1$ with radius $R$. On this background, there are three 
classes of string modes: oscillatory modes, momentum modes and 
winding modes. The momentum modes correspond to the momentum of the 
center of mass of the string, and is analogous to the corresponding 
momentum states for a point particle on $T^1$. The energies of the 
momentum modes take on the values $\frac{n}{R}$ (with positive 
integers $n$) due to usual quantization of momenta on this compact 
background. The winding modes correspond to the number of times the 
closed string winds around $T^1$. Their energies are quantized as 
$mR$, where the positive integer $m$ is the number of times the 
string winds the $T^1$. Since the string oscillatory modes have 
energies which are independent of $R$, the spectrum of the 
perturbative string is invariant under so-called T-duality: 
\be R \, \rightarrow \, \frac{1}{R}, \,\,\, (n, m) \, 
\leftrightarrow \, (m, n) \, . \ee 
The vertex operators also obey this symmetry, and by introducing 
branes T-duality can be extended to be a symmetry of the full string 
theory: it is not just a perturbative symmetry, but also a 
non-perturbative one (see e.g. \cite{Pol,Boehm}). 
 
As the temperature of the string gas increases, the number of 
accessible string oscillatory modes increase exponentially. This 
leads to the existence of a maximal temperature for a gas of strings 
in thermal equilibrium, the Hagedorn temperature $T_H$ 
\cite{Hagedorn}. This limiting temperature is reached once the 
energy density reaches string density. Because of T-duality, the 
temperature of a string gas is maximal if $R$ is equal to the string 
length $l_s$. Moreover, the higher the entropy of the string gas is 
at the duality point $R=l_s$, the larger the region of $R$ values 
for which the temperature hovers near the Hagedorn temperature. 
 
The phase during which the temperature is close to the Hagedorn 
temperature and the string gas contains a thermal distribution of 
both winding and momentum modes is called the Hagedorn phase. During 
this phase, the winding modes and the momentum modes contribute with 
opposite sign to the pressure, so that the equation of state is 
$\omega=0$ (modulo the contribution of the oscillatory modes). 
Considering the above fact, it follows from the equations of motion 
(\ref{E1}-\ref{E3}) that the string frame scale factor has a static 
solution.  The evolution of the dilaton is given by 
\be {\ddot \phi} + (d) {\dot \lambda} {\dot \phi} - 2 {\dot \phi}^2 
\, = \, - {1 \over 2} e^{\varphi} E [1 - (d) w]~ . \ee 
In the Hagedorn phase, the dilaton cannot be static, unless the 
effective field theory for the background fields is changed. In the 
radiation phase, on the other hand, the equations of motion can be 
combined to show that the dilaton approaches a constant due to the 
Hubble damping term. 
 
As discussed in \cite{Betal} (see also \cite{KKLM}), the time 
dependence of the dilaton in the Hagedorn phase which follows from 
the equations of motion for dilaton gravity conflicts with the basic 
symmetries on which string cosmology must be based. Once the dilaton 
becomes large, the string gas should include Dp branes which become 
light. The S-duality symmetry of string theory indicates that one 
should expect that the dilaton is static in the Hagedorn phase 
\cite{Betal}. In order to stabilize the dilaton, extra ingredients 
are needed.  One attempt to find a background action which is 
consistent with the idea of a Hagedorn phase in which both the scale 
factor and the dilaton are quasi-static was recently made in 
\cite{Frey}. It was based on ideas of tachyon condensation, i.e. 
introducing a moving tachyon condensate. In the context of higher 
derivative gravity actions which admit a singular bounce 
\cite{Biswas1}, it is possible to obtain a Hagedorn string gas phase 
near the bounce point in a model without a dilaton \cite{Biswas2}. 
It is still an open issue how to stabilize the dilaton more 
naturally. 
 
\begin{figure} 
\centering 
\includegraphics[totalheight=3.6in]{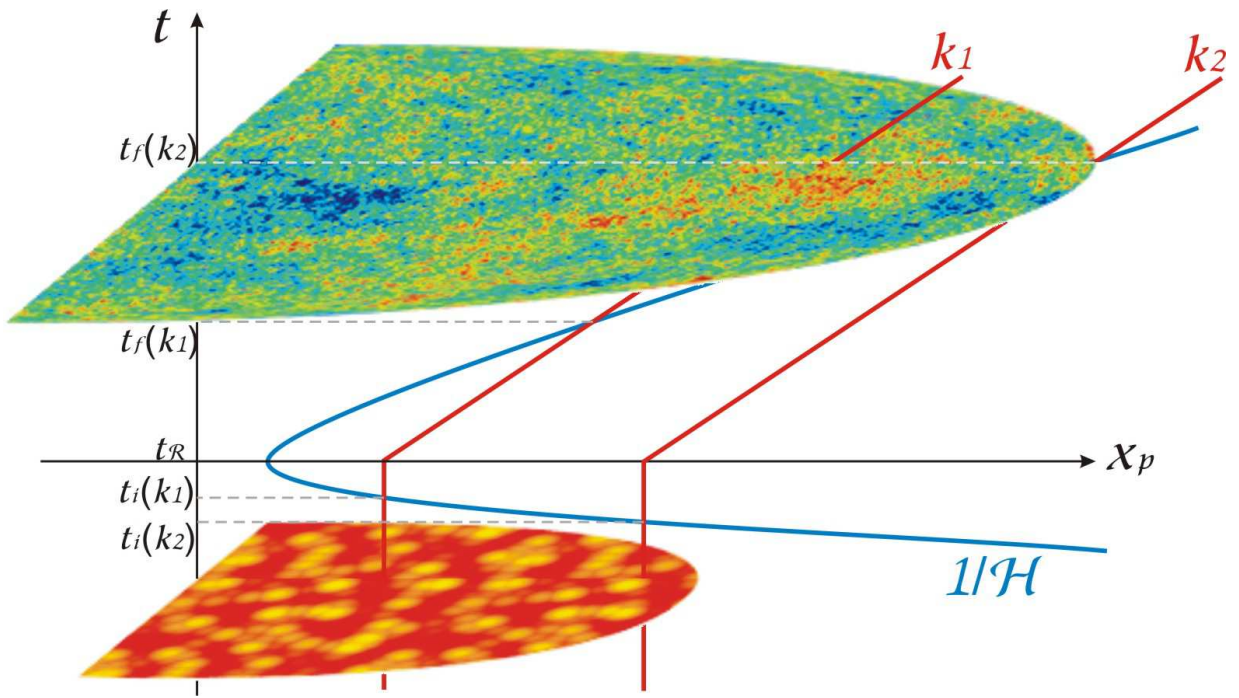} 
\caption{\small{The figure illustrates how the Hubble radius evolves 
from the Hagedorn phase to the radiation-dominated period. Different 
modes in the Hagedorn phase escape the Hubble radius with different 
physical length, and they re-enter the Hubble radius to generate 
anisotropies in the CMB. }} \label{fig:sgc} 
\end{figure}

The transition between the Hagedorn phase and the radiation phase of 
standard cosmology is smooth and is triggered by the annihilation of 
winding and anti-winding modes into string loops (which behave like 
radiation). Whereas in the Hagedorn phase the scale factor is almost 
constant, in the radiation phase it evolves as 
\be a(t) \, \sim \, t^{\frac{1}{2}} ~. \ee 
Therefore, the Hubble radius is almost infinite in the Hagedorn 
phase as shown in Fig.\ref{fig:sgc}. During the transition between 
the Hagedorn phase and the radiation phase the Hubble radius rapidly 
decreases. Once in the radiation phase, the Hubble radius increases 
linearly in time as in standard cosmology. Because of the static 
nature of the Hagedorn phase, all scales of cosmological interest 
today are sub-Hubble during this period. Provided that the Hagedorn 
phase is sufficiently long, thermal equilibrium can be established 
on these scales. 
 
Comoving scales whose physical length today corresponds to the 
current Hubble radius had a physical wavelength of the order of $1 
{\rm mm}$ at the end of the Hagedorn phase, assuming that the 
temperature at that time was of the order of the scale of Grand 
Unification. The length is constant during the Hagedorn phase since 
the universe is static. The initial conditions for fluctuations in 
SGC are very different than in inflationary cosmology. In 
inflationary cosmology, the exponential expansion of space 
red-shifts any pre-existing matter, leaving matter in a local vacuum 
state. Thus, fluctuations are quantum vacuum type. In contrast, in 
SGC matter is not red-shifted during the Hagedorn phase. If matter 
is in a thermal state, then the fluctuations will be string 
thermodynamic fluctuations. Since we are interested in length scales 
which are in the far infrared compared to the string scale, it is 
reasonable to assume that the fluctuations can be treated using 
Einstein gravity. Given this assumption, it was discovered in 
\cite{NBV} that the spectrum of resulting cosmological perturbations 
today is roughly scale-invariant with a small red tilt. The spectrum 
of gravitational waves is also almost scale-invariant, but has a 
slight blue tilt \cite{BNPV2}, yielding a distinctive signature of 
the SGC structure formation scenario for experiments. Since the 
fluctuations exit the Hubble radius and evolve outside of the Hubble 
radius as they do in inflationary cosmology, they are squeezed and 
yield acoustic oscillations in the angular power spectrum of cosmic 
microwave background anisotropies, again as in inflationary 
cosmology (see e.g. \cite{MFB} for a comprehensive survey of the 
theory of cosmological perturbations and \cite{RHBrev} for an 
overview). 
 
\section{Thermal Correlation Functions and Non-Gaussianity} 
 
In this section, we calculate the two- and three-point correlation 
functions of the scalar metric perturbations, and discuss 
implications for future experiments. 
 
At the end of the Hagedorn phase, the Hubble scale is decreasing 
from near infinity to a very small scale. After the winding modes 
annihilate with each other and create  more and more radiation, the 
universe naturally evolves into the radiation dominated phase. Since 
during the Hagedorn phase the Hubble radius is much larger than the 
physical length of scales we are interested in, we are in the 
sub-Hubble regime when matter fluctuations dominate the dynamics. 
The logic of the calculation is to compute the matter perturbations 
during the Hagedorn phase, convert them to metric fluctuations once 
the scales exit the Hubble radius at the end of the Hagedorn phase, 
and then to evolve the metric fluctuations as is usually done in 
cosmological perturbation theory. This is the logic espoused in 
\cite{NBV,BNPV1}. We will thus begin with the computation of the 
matter fluctuations. 
 
Since the Hagedorn phase is quasi-static and adiabatic, and matter 
is dominated by the thermal gas of strings, one can analyze the 
thermodynamics of these strings and calculate the thermal 
fluctuation in the standard way. Starting point for the computation 
of the thermal fluctuations is the expression for the partition 
function of a gas of closed strings on a compact space with three 
large spatial dimensions derived in \cite{Deo} (see also
\cite{Mitchell,Monica} for other work on string gas
thermodynamics). The three large 
spatial dimensions are assumed to have toroidal topology such that 
stable string winding modes exist. 
 
The correlation functions of the density fluctuation can be obtained 
from the partition function $Z$ 
\begin{equation} Z \, = \, \Sigma \exp (-\beta E_\alpha) \ , 
\end{equation} 
where $\beta=T^{-1}$. The first step of the analysis is to compute 
the average energy of a gas of closed string in a box of length $L$, 
which in the limit when the temperature $T$ is very close to the 
Hagedorn temperature is given by \cite{Ali,NBV,BNPV1} 
\begin{equation} 
\langle U \rangle \, \simeq \, \frac{L^2}{l_s^3} \ln [\frac{l_s^3 T} 
{L^2(1-\frac{T}{T_H})}]\ , 
\end{equation} 
%
where $T_H$ is the Hagedorn temperature. 
 
Then, the two-point correlation function for the fluctuation 
\be \delta \rho \, \equiv \,  \rho-\langle\rho\rangle \ee 
is given by \cite{NBV,Ali,BNPV1} 
\begin{equation}\label{rho2} 
\langle\delta\rho^2\rangle \, = \, \frac{\langle\delta 
U^2\rangle}{V^{2}} \, = \, \frac{1}{V^2}\frac{d^2 \log Z}{d\beta^2} 
\, = \, -\frac{1}{V^{2}}\frac{d\langle  U\rangle}{d\beta} \, =\, 
\frac{T}{L^4 l_s^3(1-\frac{T}{T_H})}~. 
\end{equation} 
Similarly, the three-point correlation function can be expressed as 
\cite{we} 
\begin{equation}\label{rho3} 
\langle\delta\rho^3\rangle \, = \, \frac{\langle\delta 
U^3\rangle}{V^{3}} \, = \, -\frac{1}{V^3}\frac{d^3\log Z}{d\beta^3} 
\, = \, \frac{1}{V^3}\frac{d^2\langle U\rangle}{d\beta^2} \, = \, 
\frac{T^2}{L^7 l_s^3 (1-\frac{T}{T_H})^2}~. 
\end{equation} 
%
Note that the linear density fluctuations is highly gauge-dependent
on super-Hubble scales. Here, however, we are only following the
matter perturbations on scales smaller than the Hubble radius. On
these scales, the gauge-dependence is negligible.
 
In the transition period between the Hagedorn phase and the 
radiation phase, the Hubble radius decreases fast, but the physical 
length of the fluctuations is almost unchanging. Once the 
fluctuations leave  the decreasing Hubble radius, the fluctuations 
freeze out. During the radiation phase, the perturbations red-shift 
as standing waves, but maintain the information about their thermal 
origin in the initial Hagedorn phase. Since the fluctuations 
re-enter the Hubble radius at late times as coherent standing waves, 
they will induce acoustic oscillations in the angular power spectrum 
of CMB anisotropies in the same way as happens in inflationary 
cosmology. 
 
The density fluctuations $\delta \rho_{\bf k}$ in momentum space are 
given by those in position space via 
\begin{equation}\label{rhok} 
  \delta\rho_{\bf k} \, = \, k^{-{3\over2}}\delta\rho~. 
\end{equation} 

When scales exit the Hubble radius, we determine the induced metric 
perturbation using the Einstein constraint equations (we are making 
the assumption that the fluctuations on the infrared scales relevant 
to current observations are indeed described by the Einstein 
equations). As shown in detail in \cite{BNPV1} 
the scalar metric fluctuation $\Phi$ is determined via 
\begin{equation}\label{Phik} 
  \Phi_{{\bf k}} \, \sim \, 4\pi G \delta\rho_{\bf k}(\frac{a}{k})^{2}~, 
\end{equation} 
where $G$ is Newton's constant, and $\Phi_{{\bf k}}$ is the Fourier 
mode of the longitudinal gauge metric perturbation defined via the 
following expression for the metric including scalar metric 
inhomogeneities (in the absence of anisotropic stress): 
\be ds^2 \, = \, a^2\left( -(1+2\Phi)d\eta^2+(1-2\Phi)d{\bf x}^2 
\right) \, . \ee 
Note that in the gauge we are using, on super-Hubble scales the
fluctuations in $\Phi$ and $\delta \rho$ have the same spectrum,
the factor $k^2$ in (\ref{Phik}) being replaced by the square of
the Hubble paramter ${\cal H}$
in conformal time. However, at Hubble radius crossing
${\cal H} = k$, and thus we obtain the Poisson-like equation
(\ref{Phik}).
  
Since the equation of state parameter $1 + w$ (where $w$ is the 
ratio of pressure to energy density) does not change substantially, 
the amplitude of the metric perturbation is almost constant after it 
leaves the Hubble radius. Thus, we can calculate the late-time power 
spectrum of cosmological fluctuations from the Hagedorn phase matter 
perturbations making use of the constraint equation (\ref{Phik}) and 
the Hubble radius crossing condition $k/a(t_H(k)) = H(t_H(k))^{-1}$, 
yielding 
\begin{equation} 
\langle \Phi_{\bf k}^2 \rangle \, \sim \, (4\pi G)^2\frac{ 
T}{l_s^3(1-\frac{T}{T_H})}k^{-3} ~, 
\end{equation} 
where $t_H(k)$ is the time when the scale $k$ exits the Hubble 
radius during the transformation period. 
 
Therefore the power spectrum of $\Phi_{\bf k}$ can be expressed as 
\begin{equation}\label{PPhi} 
 P_{\Phi} \, \equiv \, \frac{k^3}{2\pi^2} \langle\Phi_{\bf k}^2\rangle 
 \, \simeq \, 8G^2 \frac{ T}{l_s^3(1-\frac{T}{T_H})} \, \simeq \, 
  8(\frac{l_p}{l_s})^4 \frac{ 1}{(1-\frac{T}{T_H})}~, 
\end{equation} 
where $l_p$ is the Planck length, and the temperature $T$ is to be 
evaluated at the time $t_H(k)$. 
 
Similarly, the three-point correlation function of $\Phi_{\bf k}$ 
after leaving thermal equilibrium is expressed as 
\begin{equation} 
\langle \Phi_{\bf k}^3 \rangle \, \sim \, (4\pi G)^3 \frac{  T^2 
H(t_H(k))}{ l_s^3(1-\frac{T}{T_H})^2}  k^{-9/2}~. 
\end{equation}

When the physical wavelength of the mode $k$ is much larger than the 
Hubble radius, the usual practise in cosmological perturbation 
theory is to focus on the conserved quantity ${\cal R}_{\bf k}$, the 
curvature fluctuation in co-moving gauge, given by 
\begin{equation} {\cal R} \, \simeq \, \zeta \, = \, 
\Phi + \frac{2}{3}\frac{H^{-1}\dot \Phi+\Phi}{1 + w}~. 
\end{equation} 
The constancy of ${\cal R}$ is then used to relate early time to 
late time fluctuations. Since in the Hagedorn phase $w = 0$ and in 
the radiation dominated universe $w = 1/3$, then, up to a order one 
constant, the constancy of ${\cal R}$ implies the constancy of 
$\Phi$, the variable which determines the CMB anisotropies. 
 
Thus, we can directly compute the non-Gaussianity estimator $f_{\rm 
NL}$, and it takes the form 
\begin{equation}\label{ng} 
f_{\rm NL} \, \sim \, k^{-\frac{3}{2}}\frac{\langle {\cal R}_{\bf 
k}^3 \rangle} 
  {\langle {\cal R}_{\bf k}^2 \rangle \langle {\cal R}_{\bf k}^2 
  \rangle}\sim \frac{l_s^3 H(t_H(k))} {4\pi l_p^2 }~. 
\end{equation} 

Using the condition of modes escaping and re-entering the Hubble 
radius $k=a(t_H(k)) H(t_H(k)) = a H$ and $k_0 = a_0 H_0$, and 
considering the relationship between scale factor and temperature, 
we can estimate the value of the non-Gaussianity to be 
\begin{equation} 
f_{\rm NL} \, \simeq \, (\frac{l_s} 
  {l_p})^2 \frac{H}{T} \, = \, (\frac{l_s} 
  {l_p})^2  \frac{H_0}{T_0} \frac{k}{k_0} \, \simeq \, 
(\frac{l_s}  {l_p})^2 \times 10^{-30} \frac{k}{k_0} ~, 
\end{equation} 
where the subscript $0$ on $H$,$T$, and $k$ represent today's 
values, subscript $t$ represents the values of Hagedorn phase. 
 
From the above relation, we see that the non-Gaussianity estimator 
$f_{NL}$ depends linearly on the mode $k$. This reflects the fact 
that thermal string fluctuations lead to non-Gaussianities which are 
of order unity on microscopic scales, but which are suppressed on 
larger scales in accordance to the Central Limit Theorem. This means 
that modes reentering the Hubble radius earlier have a larger 
non-Gaussianity. This is very different from what happens in 
inflationary models, where the non-Gaussianities are almost scale 
invariant. 
 
Moreover, $f_{NL}$ depends sensitively on the string scale . If the 
string scale is high (comparable to the Planck scale or the scale of 
Grand Unification), then $f_{NL}$ will be be too small to be 
observed in the future experiments. However, if the string scale is 
at the TeV scale, then the non-Gaussianity on a scale of $k_0$ is 
$f_{\rm NL}\sim {\cal O} (1)$. For such a string scale, however, it 
requires fine tuning of the temperature $T$ of the string gas in the 
Hagedorn phase to obtain a power spectrum with reasonable amplitude. 
Therefore if the future experiment obtains a non-vanishing 
non-Gaussianity, this would pose a significant challenge to string 
gas cosmology. On the other hand, if non-Gaussianity with an 
amplitude growing linearly with $k$ were to be detected, it would 
provide a strong signature in support of string gas cosmology with a 
low string scale.

\section{Conclusion and Discussion} 
 
In this paper, we have calculated the non-Gaussianity parameter 
$f_{NL}$ in string gas  cosmology using the formalism developed in 
\cite{we}. We obtain a result which grows linearly with $k$. For a 
high string scale, the amplitude of the predicted non-Gaussianities, 
however, is much too small to be observable on cosmological scales 
today. 
 
In general inflation models, the shape of the non-Gaussianity 
\cite{NG} is almost scale invariant, since during the inflation 
period different modes exit the Hubble radius in almost the same 
physical environment (the time-translation invariance of the De 
Sitter phase). In contrast, in SGC the fluctuations exit the Hubble 
radius in a phase in which the string gas is rapidly changing its 
character. Although the holographic scaling of the specific heat of 
a closed string gas can ensure the scale-invariance of the power 
spectrum, the three-point correlation function cannot be 
scale-invariant. 
 
If the string scale is high, SGC cannot give a large 
non-Gaussianity. The amplitude could be very hard if not completely 
impossible to detect in future experiments.  If the string scale is 
decreased to the TeV scale, which could happen in some scenarios 
with warped geometry, then the non-Gaussianity could be large. 
However, in this one would require a fine-tuning on the temperature 
$T$ of the thermal string gas in order to obtain a power spectrum 
with reasonable amplitude. 
 
We should also emphasize that the method used here to calculate the 
three-point correlation function in SGC is general. It can be 
applied to inflationary scenarios with thermal fluctuations. The 
very early universe has many thermal elements, whose fluctuations 
could be the source of non-Gaussianities in the CMB. In the case of 
inflation, the physical wavelength of fluctuations when they exit 
the Hubble radius is microscopic, compared to the situation in SGC 
where the length is macroscopic. This leads to a much larger 
amplitude of the non-Gaussianities in thermal inflation models, as 
shown in \cite{we}. 
 
We wish to mention a further caveat to our analysis: we have 
neglected the possible existence of cosmic superstrings 
\cite{Witten} as a remnant of the early Hagedorn phase. Such strings 
would induce extra contributions to the spectrum of fluctuations and 
would lead to non-Gaussianities. 
 
\section*{Acknowledgments} 
 
We would like to thank Miao Li and Bo-Qiang Ma for discussion. One 
of us (RB) wishes to acknowledge the hospitality of the KITPC during 
a visit during which this project was started. We are grateful to 
KITPC for its wonderful programme on String and Cosmology, in which 
much of this work was done and discussed. BC would like to thank OCU 
for its hospitality, where this project was finished. The work was 
partially supported by NSFC Grant No. 10535060, 10775002. RB is 
supported in part by funds from NSERC, the Canada Research Chair 
program and a FQRNT Team Grant.

\end{document}